# Ultrafast multi-photon excitation of $ScVO_4:Bi^{3+}$ for luminescence thermometry

DAVID ESCOFET-MARTIN[1,2,3], ANTHONY O. OJO[2,*], AND BRIAN PETERSON[2]

[1]These authors contributed equally to this work
[2]Institute for Multiscale Thermofluids, School of Engineering, The University of Edinburgh, Edinburgh EH9 3FD, United Kingdom
[3]Email: david.escofet@ed.ac.uk
*Corresponding Author: anthony.ojo@ed.ac.uk

**Abstract:** *We demonstrate a multi-photon excitation (MPE) scheme for luminescence thermometry using $ScVO_4:Bi^{3+}$. MPE is performed using a 37 fs Ti:Sapphire laser pulse centred at 800 nm. Log-log plots of the phosphorescence intensity versus excitation power show that the 800 nm MPE of $ScVO_4:Bi^{3+}$ involves a 2- and 3-photon absorption process in comparison to a single-photon excitation (SPE) process at 266 nm and 400 nm. Spectroscopic investigation shows that with the 800 nm MPE and 266 nm SPE schemes, the emission spectra of $ScVO_4:Bi^{3+}$ are similarly characterized by emissions of the $VO_4^{3-}$ groups and $Bi^{3+}$. The MPE is advantageous to suppress fluorescence which interfere with the phosphorescence signal. We demonstrate this aspect for a $ScVO_4:Bi^{3+}$ coating applied on an alumina substrate. The luminescence lifetime is calibrated with temperature over 294-334 K; the MPE scheme has an equally impressive temperature sensitivity (3.4-1.7% / K) and precision (0.2-0.7K) compared to the SPE schemes. The MPE scheme can be applied to a variety of phosphors and is valuable for precise temperature measurements even in applications where isolating interfering background emissions is challenging.*

http://doi.org/10.1364/OL.445763

Measurements of temperature are vital in describing physical, biological, and chemical processes. Non-contact optical techniques for measuring temperature, such as phosphor thermometry, allow us to monitor temperature and understand heat transfer processes in a variety of engineering [1-4] and biological applications [5, 6]. With phosphor thermometry, the temperature-dependent luminescent properties of inorganic materials called thermographic phosphors (TGPs) are exploited to infer temperature. Pulsed lasers are often used to provide an instantaneous excitation radiation to drive the photoluminescence (PL) process in TGPs. Most TGPs feature a broad excitation spectrum in the UV [1], allowing the use of widely available pulsed ns lasers emitting UV radiation for thermometry.

Pulsed excitation schemes deployed for thermometry with ns UV lasers largely promote single-photon excitation (SPE). Upconversion in TGPs through multiple metastable states with good excitation efficiency are possible with near-infrared (NIR) ns lasers [6-8]. However, TGPs that rely on upconversion often feature sensitizers and activators as dopants within the TGP material to allow for the intermediate metastable states [9].

The development of chirped pulse amplification [10], and the evolution of commercially available ultra-short laser systems has led to technological breakthroughs that positively affected our society. Thus, high energy femtosecond (fs) lasers are broadly available and have found use in a range of applications as they provide negligible thermal interaction with a variety of materials [11, 12]. These fs lasers often provide NIR radiation with peak powers in order of terawatts compared to megawatts from ns lasers. NIR excitation with terawatt laser pulses allows to efficiently excite electronic transitions via a multi-photon excitation (MPE) process [13]. These electronic transitions are driven through virtual states which have negligible lifetimes [14].

For thermometry in engineering applications involving the use of liquid fuels, or coatings on ceramic substrates, SPE can lead to the generation of fluorescence signals that could interfere with TGP measurements. The same holds true in biomedical applications where auto-fluorescence of biological parts can occur [6]. Infrared MPE of TGPs generally provides the opportunity to avoid such unwanted background fluorescence signals. Furthermore, because MPE does not require electronic transitions via intermediate metastable states, it can be applied to a wide range of TGPs which do not have these intermediate states. While MPE suffers from absorption cross-sections that are orders of magnitude below the equivalent SPE absorption cross-section, the use of fs lasers for MPE mitigates the reduced excitation efficiency by providing high peak power.

This work exploits the high peak power of a fs laser for thermometry through the MPE of bismuth-doped scandium vanadate ($ScVO_4:Bi^{3+}$). We employ $ScVO_4:Bi^{3+}$ due to the advantages it offers for thermometry. Over the 20-60°C temperature range, $ScVO_4:Bi^{3+}$ was shown to offer a lifetime sensitivity of 2.2%/°C, which is among the highest reported for lifetime-based thermometers, yielding sub-Kelvin temperature precision [15]. Furthermore, its short luminescence lifetime (< 10 μm at room temperature [15, 16] makes it promising for kHz-rates measurements. Hence, the objectives of this work are to; (1) investigate and demonstrate the MPE of $ScVO_4:Bi^{3+}$ using an 800 nm fs laser, (2) demonstrate the utility of the MPE scheme in suppressing fluorescence, and (3) evaluate the temperature sensitivity and precision obtainable using the MPE scheme.

The TGP, bismuth-doped (1 mol%) scandium vanadate $ScVO_4:Bi^{3+}$, used in this work was obtained commercially (Phosphor Technology, UK). Figure 1 shows the measured X-ray diffraction (XRD) pattern for the $ScVO_4:Bi^{3+}$ sample. For comparison, the $ScVO_4:Bi^{3+}$ XRD pattern is shown together with the XRD data set for un-doped $ScVO_4:Bi^{3+}$ (ICSD No. 78073/CSD No. 1638403 [17]) retrieved from the Cambridge Structural Database [18]. The diffraction peaks of the $ScVO_4:Bi^{3+}$ sample are comparable to the ICSD data set and to previously reported data [19], indicating that the sample crystallizes in the tetrahedral zircon-type space group $I4_1/amd$.

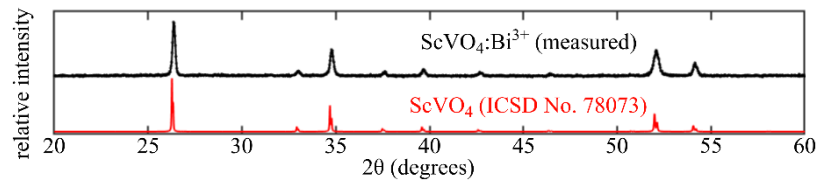

Fig. 1. $ScVO_4:Bi^{3+}$ and reference ScVO4 XRD patterns.

Ultrafast MPE of $ScVO_4:Bi^{3+}$ is performed using a near Fourier transformed 37 fs Ti:Sapphire laser pulse centred at 800 nm. A Frequency-Resolved Optical Gating device was used to characterise the pulse duration of the 800 nm beam (9 mm gaussian ($1/e^2$)). Figure 2a shows the spectrum of the 800 nm radiation with a ≈ 22 nm FWHM. The MPE of $ScVO_4:Bi^{3+}$ is compared with SPE of $ScVO_4:Bi^{3+}$ under two excitation schemes. The first is the 2nd harmonic (400 nm, < 1 ps laser pulse, 3.6 mm gaussian ($1/e^2$)) generated by doubling the 800 nm beam (37 fs laser pulse) from the Ti:Sapphire laser using a BBO crystal. The other SPE scheme is performed using the 4th harmonic (266 nm, 10 ns laser pulse, 3 x 8 $mm^2$) of an Nd:YAG laser. Each lasers operated at 1 kHz. The laser average power used for the 800 nm MPE, 400 nm and 266 nm SPE schemes was in the range 0.5-3.0 W (0.5-3 mJ/pulse), 0.02-0.4 W (0.02-0.4 mJ/pulse) and 0.03-0.1 W (0.03-0.1 mJ/pulse), respectively. A photomuliplier tube, equipped with a 607±18 nm filter, detected the temporally evolving phosphorescence signals. Emission spectra, including the entire phosphorescence decay, were detected using a sCMOS KURO camera (0.05 nm/pixel, 0.3 nm resolution) mounted on a HRS-750 spectrometer (300 grooves/mm grating, 70 μm entrance slit). The spectrometer was wavelength- and intensity-calibrated with calibration lamps. Phosphorescence from $ScVO_4:Bi^{3+}$ was imaged into the spectrometer entrance slit using a 100 mm plano-convex lens providing a ≈ 1:1 magnification. A 450 nm longpass filter and a 750 nm shortpass filter were used during the 400 nm SPE and 800 nm MPE schemes.

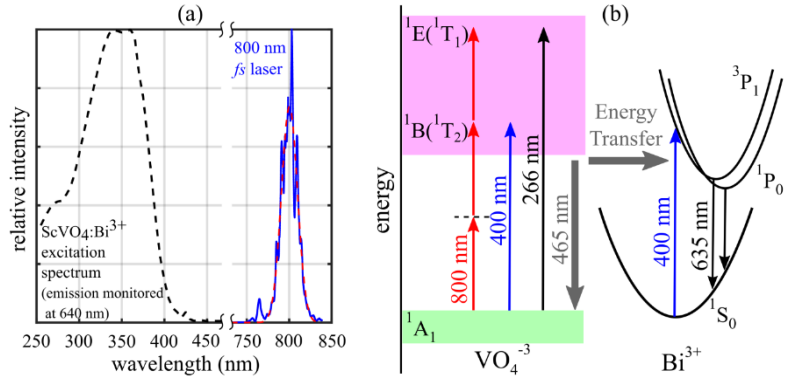

Fig. 2. (a) Excitation spectrum of ScVO$_4$:Bi$^{3+}$ (data plotted with permission of Phosphor Technology, UK) and spectrum for the fs 800 nm laser light (red-dash line is a fit), (b) Schematic configuration coordinate diagram for ScVO$_4$:Bi$^{3+}$.

Figure 2b shows the configuration coordinate diagram for PL of ScVO$_4$:Bi$^{3+}$. Energy transitions accompanying the excitation schemes are shown in Fig. 2b. The photon energies for the 400 nm and 266 nm SPE are equivalent to the 800 nm 2-photon and 3-photon MPE energies. Upon excitation of ScVO$_4$:Bi$^{3+}$ by UV (e.g. 266 nm radiation), electrons are promoted from the $^1A_1$ state to $^1E(^1T_1)$ from where they relax to $^1B(^1T_2)$ in the VO$_4^{3-}$ groups of the vanadate host. The $^1B(^1T_2)$ → $^1A_1$ transition produces a blue broad emission at ≈ 465 nm. However, because the absorption band of Bi$^{3+}$ doped in ScVO$_4$ (extending to 430 nm (see Fig. 2a)) overlaps with part of the broad VO$_4^{3-}$ emission, the energy absorbed by the VO$_4^{3-}$ groups can be transferred to Bi$^{3+}$ through resonant cross relaxation [16, 20]. Thus, $^1S_0$ → $^3P_1$ electronic transitions can occur in Bi$^{3+}$. With a 400 nm excitation scheme, absorption within 360-430 nm for ScVO$_4$:Bi$^{3+}$ is mainly from $^1S_0$ → $^3P_1$ of Bi$^{3+}$ [16]. The red broad emission (at ≈ 635 nm) of ScVO$_4$:Bi$^{3+}$ is attributed to the $^3P_1$ → $^1S_0$ and $^3P_0$ → $^1S_0$ transitions of Bi$^{3+}$. It is important to note that the Bi$^{3+}$ doping concentration in the ScVO$_4$ host matrix influences the PL processes in ScVO$_4$:Bi$^{3+}$ [16, 20]. With doping concentration < 1 mol%, energy transfer from the VO$_4^{3-}$ groups to the Bi$^{3+}$ is less efficient so that emissions from the VO$_4^{3-}$ groups (465 nm) and Bi$^{3+}$ (635 nm) are evident. Doping concentration of 1 mol% presents the most efficient energy transfer from the host to the dopant, with the emission spectrum featuring mainly Bi$^{3+}$ (635 nm) emission.

The excitation spectrum of ScVO$_4$:Bi$^{3+}$ in Fig. 2a shows that there is no absorption band above 430 nm, such that the TGP does not support excitation above 430 nm [16]. With NIR (e.g. 800 nm) excitation, non-linear optical processes such as upconversion through multiple metastable excited states [8], or virtual states could occur, with the former having higher probability than the latter. However, intermediate metastable excited states have not been reported for the VO$_4^{3-}$ groups and Bi$^{3+}$. Thus, the upconversion process will only occur through virtual states. Hence, we describe the non-linear optical absorption at 800 nm in relation to MPE processes which occurs through intermediate virtual state in the energy gap in ScVO$_4$:Bi$^{3+}$. As shown in Fig. 2b, with the illumination of ScVO$_4$:Bi$^{3+}$ at 800 nm, there could be simultaneous absorption of two photons which enhances the $^1S_0$ → $^3P_1$ transition of Bi$^{3+}$. In addition, due to the high photon flux, a 3-photon excitation process can occur simultaneously with the 2-photon excitation process, so that electrons are excited into the charge transfer state of the VO$_4^{3-}$ groups. To our knowledge, we present the first demonstration of MPE of a TGP which relies on energy transitions via a virtual excited state for potential application in temperature sensing.

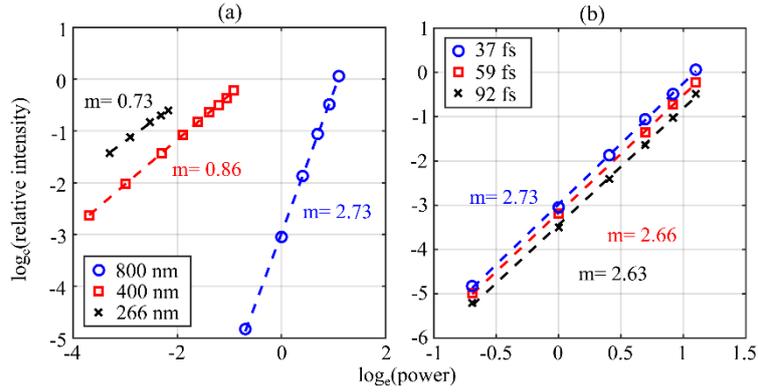

Fig. 3. Natural log-log plots of phosphorescence intensity versus excitation average power, (a) for each scheme, (b) for different pulse durations (FWHM) with 800 nm excitation.

To investigate the ultrafast MPE of $ScVO_4:Bi^{3+}$ in comparison to its SPE, the phosphorescence intensities were measured as a function of the average excitation power at room temperature. These measurements were performed during the proposed MPE (800 nm) and the SPE (266 nm and 400 nm) of $ScVO_4:Bi^{3+}$ coated on an aluminium metal bar. Figure 3 shows the natural log-log plots of the initial phosphorescence intensity, recorded with the PMT, versus excitation power. Figure 3a shows a plot of the intensity-vs-power dependence under the MPE for $ScVO_4:Bi^{3+}$ with the slope being equal to 2.73. This value, near 3, indicates that the non-linear optical absorption process associated with excitation of $ScVO_4:Bi^{3+}$ at 800 nm is dominated by a 3-photon absorption to yield the phosphorescence. For the SPE schemes at 400 nm and 266 nm excitation, Fig. 3a shows the measured slopes to be 0.83 and 0.73, respectively. This value, near unity, confirms a SPE process for these schemes. Deviations from a cubic or linear intensity-vs-power dependence for the respective measured slopes (< 3 for the MPE and < 1 for SPE) is potentially due to inter-system crossing to a non-phosphorescent state. In addition, the deviation of slope from the cubic dependence for the MPE could also be due to excited state absorption. However, we expect that with the MPE and SPE of $ScVO_4:Bi^{3+}$, electronic transitions to the same initial excited states of $Bi^{3+}$ would occur such that inter-system crossing processes would be similar regardless of the excitation scheme employed [7].

We further examined the intensity-vs-power dependence using two additional pulse durations, 59 fs and 92 fs, for MPE of $ScVO_4:Bi^{3+}$. The pulse duration is adjusted by tweaking the fs laser compressor, inducing chirp in the laser pulses while maintaining the output energy. Figure 3b shows the intensity dependence when using the 37, 59 and 92 fs pulses for MPE. The measured slope of these data set is equal to 2.73, 2.66 and 2.63, respectively. The slope decreases slightly with increasing excitation pulse duration, however, results suggest that the 800 nm MPE of $ScVO_4:Bi^{3+}$ is dominated by 3-photon absorption process. Due to the higher instantaneous power, the relative phosphorescence intensity from the 37 fs pulse is, on average, 32% and 66% greater than the 59 fs and 92 fs pulses, respectively. This indicates an increased excitation efficiency for the shortest pulse. To obtain similar phosphorescence intensity with the different excitation schemes ($\log_e$(intensity)=-1), the 800 nm MPE requires 2 W, while the 400 nm and 266 nm SPE require 0.17 W and 0.07 W, respectively. The excitation efficiency for MPE could be improved by focusing the fs beam when exciting the TGP.

Figure 4a shows the room temperature emission spectra of $ScVO_4:Bi^{3+}$ under the 800 nm MPE compared with the spectra under the 266 nm and 400 nm SPE schemes. The dominant feature of the $ScVO_4:Bi^{3+}$ spectra under the MPE and SPE schemes is the broad emission of $Bi^{3+}$ with peak at $\approx$ 630 nm. This suggests that electronic transitions occur to the same initial excited states in $Bi^{3+}$ regardless of the excitation scheme used. However, a clear difference in the spectra for these excitation schemes is in the contribution of the vanadate host ($VO_4^{3-}$ groups) emission, which peaks at $\approx$ 460 nm. The presence of the host and dopant emissions suggests that our TGP sample exhibits features of a TGP with dopant concentration < 1 mol% [16].

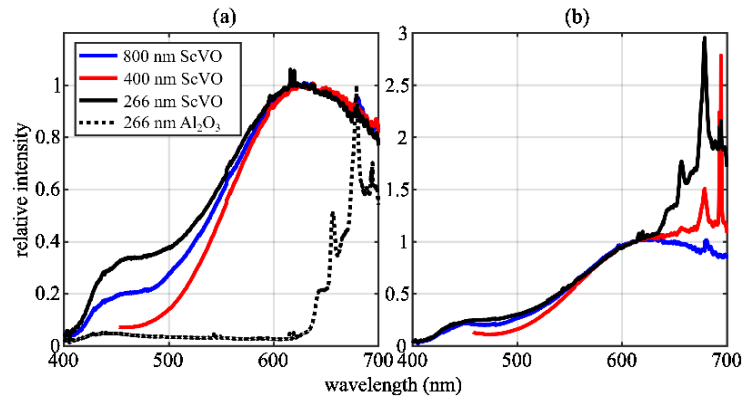

Fig. 4. Normalized emission spectra (0.3 nm resolution) of (a) $ScVO_4:Bi^{3+}$ (under the MPE and SPE schemes) and alumina, (b) $ScVO_4:Bi^{3+}$ coated on an alumina substrate.

The $VO_4^{3-}$ group emission is significantly weaker in the spectrum for the 400 nm SPE since electronic transitions with this excitation scheme mainly occurs from $^1S_0 \rightarrow {}^3P_1$ of $Bi^{3+}$. The $ScVO_4:Bi^{3+}$ spectra for the 800 nm MPE and 266 nm SPE features contributions from the $VO_4^{3-}$ group emission at ≈ 460 nm. The relative intensity ratio of $VO_4^{3-}$ to $Bi^{3+}$ emission from $ScVO_4:Bi^{3+}$ under 800 nm MPE is weaker (≈ 0.2) than for the relative intensity ratio of these emissions (≈ 0.35) under 266 nm SPE. For the 800 nm 3-photon excitation, like with the 266 nm SPE, electrons are excited into the charge transfer states of $VO_4^{3-}$. In addition, 2-photon absorption simultaneously occur for the 800 nm MPE, driving the $^1S_0 \rightarrow {}^3P_1$ transitions of $Bi^{3+}$. In Fig 4, the origin of peaks at 616 nm, 620 nm and 680 nm is being investigated, however, they do not affect our conclusions here.

Having established the MPE of $ScVO_4:Bi^{3+}$, we demonstrate the utility of this scheme in suppressing background emissions. Such background emissions are evident in applications where a TGP is coated on a substrate that could fluoresce. Abrasion of the coating due to harsh conditions would expose the substrate to laser radiation, and lead to emission which interferes with phosphorescence signals. Here, by applying a thin coating of $ScVO_4:Bi^{3+}$ (2 x 12 mm$^2$) on a ceramic alumina ($Al_2O_3$) substrate, we probed the coating with the SPE and MPE laser beams. These laser beams illuminated both the coating and the alumina substrate. Figure 4a shows the fluorescence from the uncoated alumina material under 266 nm excitation, appearing as the dash-line plot. There, the high-intensity spectral features of alumina emission are observed at wavelengths above 600 nm. Figure 4b shows the emission spectra from the $ScVO_4:Bi^{3+}$-coated alumina material under the MPE and SPE schemes. As shown, the interference of alumina emission on the phosphorescence emission is evident with both 400 nm and 266 nm SPE schemes. Such interference is absent with the 800 nm MPE approach. For thermometry with the lifetime-based approach, a fitting window can be set to avoid the fast fluorescence signals. However, in the spectral intensity ratio approach using fast TGPs, or with the lifetime dual-frame ratio-based method [15], it is difficult to isolate interfering fluorescence signals. MPE strategies can be implemented in such scenarios to suppress fluorescence emissions that leak through optical filters.

We then investigate the temperature dependence and sensitivity of the luminescence lifetime of $ScVO_4:Bi^{3+}$ under the 800 nm MPE scheme. The results are compared with those under the 266 nm and 400 nm SPE schemes. We do this by probing a heated TGP-coated aluminum metal bar equipped with a thermocouple and thermal insulation, which cooled to room temperature. Here, we only evaluate the luminescence decay time as a quantity to infer temperature without reference to photo-physical models describing the luminescence decay kinetics. Thus, luminescence decay signals, sampled at 25 MS/s, were approximated by a mono-exponential decay; decay times were evaluated using an iterative algorithm [21].

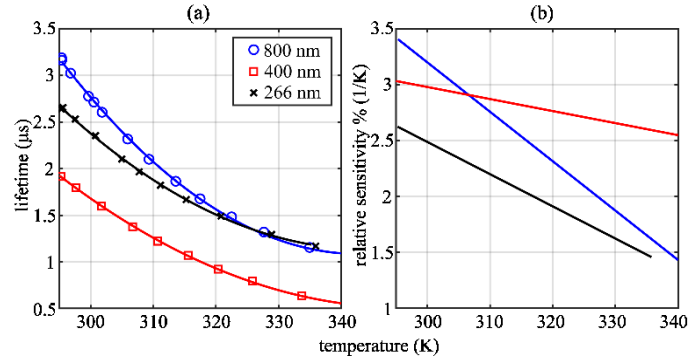

Fig. 5. (a) ScVO$_4$:Bi$^{3+}$ lifetime vs. temperature. Solid lines represent a data fit. (b) Relative temperature sensitivity vs. evaluation from the respective fit.

Figure 5a shows the luminescence lifetime temperature dependence under the MPE and both SPE schemes. The luminescence lifetime is similar for the 800 nm and 266 nm, and ≈ 0.7 μm lower for the 400 nm SPE. As earlier discussed, these excitation schemes induce different energy transfer processes, such as the charge transfer from the VO$_4^{3-}$ groups to Bi$^{3+}$ ions evident in the 800 nm MPE and 266 nm SPE, and the direct excitation of the Bi$^{3+}$ ions under 400 nm SPE. The mechanism behind the luminescence lifetimes under these excitation schemes requires further investigation. However, in this work, due to thermal quenching, the luminescence lifetimes decrease monotonically with temperature under these excitation schemes. It generally follows that the fast lifetime (< 4 μs) of ScVO$_4$:Bi$^{3+}$ under 800 nm MPE also makes it suitable for kHz-rates thermometry.

Figure 5n shows the relative temperature sensitivity, $|(1/\tau)(d\tau/dT)|$, assessed for the MPE and SPE schemes. At 294 K, the temperature sensitivity is 3.4% / K, decreasing to 1.7% / K at 334 K for the MPE scheme. Within the 294-340 K temperature range, on average, the temperature sensitivity from the 800 MPE is 2.5% / K compared with 2.8% / K for 400 nm SPE and 2.1% / K for 266 nm SPE. Furthermore, with the MPE scheme, a temperature precision of 0.2 K and 0.7 K is obtained at 294 K and 334 K, respectively. The respective temperature precision obtained at similar temperatures under the 400 nm SPE is 0.3 K and 0.7 K, and under 266 nm SPE is 0.3 K and 0.6 K. Thus, the temperature sensitivity and precision attainable with the MPE of ScVO$_4$:Bi$^{3+}$ is generally comparable with those under the SPE schemes.

In summary, we have demonstrated the MPE of ScVO$_4$:Bi$^{3+}$ using an 800 nm fs laser. The phosphorescence intensity dependence on average excitation power under the MPE and two SPE schemes was compared to establish the non-linear optical absorption process involved. We show that the MPE involves a 2- and 3- photon process. We confirmed the MPE process via the similarities in the emission spectrum of ScVO$_4$:Bi$^{3+}$ under 266 nm SPE and under the 800 nm MPE. The utility of the MPE scheme in suppressing fluorescence interference was demonstrated. Finally, the thermometric capabilities of the TGP under the MPE scheme are demonstrated within the temperature range of 293 - 340 K. The temperature sensitivity and precision is comparable with those obtainable with conventional SPE schemes at 266 nm and 400 nm. As short-pulse NIR lasers are becoming available, the MPE scheme can be applied to a variety of TGPs and is valuable where it is difficult to isolate interfering background emissions particularly for temperature imaging.

**Funding.** Engineering and Physical Sciences Research Council (EP/V003283/1, EP/P020593/1); European Research Council (759546).

**Acknowledgements.** We thank Dr. Benoit Fond for discussions and Dr. Nicholas Odling for the XRD measurements.

**Disclosures.** The authors declare no conflicts of interest.

**Data Availability.** Data underlying the results presented in this Letter are not publicly available at this time but may be obtained from authors upon reasonable request.